\documentclass[12 pt, letterpaper]{article}
\usepackage{mathrsfs}
\usepackage{amsmath}
\usepackage{amssymb, amsthm}
\usepackage{graphicx}
\usepackage{color}
\usepackage{pstricks}
\usepackage{tikz}
\usepackage{subcaption}
\usepackage[margin=0.9in,letterpaper]{geometry}
\usepackage{url}

\newcommand{\R}{{\mathbb{R}}}

\newcommand{\Hb}{{\mathbb{H}}}
\newcommand{\Sb}{{\mathbb{S}}}

\newtheorem{theorem}{Theorem}
\numberwithin{theorem}{section}
\newtheorem{lemma}[theorem]{Lemma}
\theoremstyle{definition}
\newtheorem{example}[theorem]{Example}
\newtheorem{definition}[theorem]{Definition}
\newtheorem{conjecture}[theorem]{Conjecture}

\newtheorem{problem}{Problem}
\usepackage{titlesec}
\titleformat{\section}{\large\bfseries}{\thesection}{1em}{}
\everymath{\displaystyle}
\title{Identify Equivalent Frames}
\author{Xuemei Chen\thanks{Department of Mathematical Sciences, New Mexico State University, 88011 NM, USA} \and Yang Chu\thanks{Undergraduate student in Mathematics, University of San Francisco, 94117 CA, USA} \and Min Zheng\thanks{Master student in Statistics, Columbia University, 10027 NY, USA}}
\begin{document}

\maketitle
\begin{abstract}
A frame is an overcomplete set that can represent vectors/signals faithfully and stably. Two frames are equivalent if signals can be essentially represented in the same way, which means two frames differ by a permutation, sign change or orthogonal transformation. Since these operations are combinatorial in nature, it is infeasible to check whether two frames are equivalent by exhaustive search. In this note, we present an algorithm that can check this equivalence in polynomial time. Theoretical guarantees are provided for special cases.
\end{abstract}

\section{Introduction}

A frame for a Hilbert space $\Hb$ is a sequence of vectors $\{f_i\}_{i\in I}\subset\Hb$ for which there exist constants $0 < A \leq B < \infty $ such that for every $x \in  \Hb$, 
\begin{equation}\label{equ:frame}
A{\|x\|}^2 \leq \sum_{i}{| \langle x,f_i\rangle |}^2 \leq B{\|x\|}^2.
\end{equation}
A frame is called \emph{tight} if $A=B$. Furthermore, a frame is \emph{Parseval} if $A = B = 1$.

This paper will focus on frames in the Euclidean space $\R^n$, so $\|\cdot\|$ denotes the Euclidean norm. It is easy to show that a finite collection of vectors $F=\{f_i\}_{i=1}^k$ is a frame of $\R^n$ if and only if they span $\R^n$.  We will abuse the notation and use $F$ for the matrix $[f_1, f_2, \cdots, f_k]$ when appropriate.

A frame is a generalization of a basis with the flexibility of redundancy, which provides stability and robustness. It has found numerous applications in signal processing~\cite{RV91}, coding theory~\cite{SH03, HP04}, imaging~\cite{CS10}, and data processing in general.

By definition, the middle term in \eqref{equ:frame} is essential in shaping up the frame. We have 
$$\sum_{i=1}^k{| \langle x,f_i\rangle |}^2= {\|F^Tx\|}^2= \langle F^Tx,F^Tx \rangle= x^TFF^Tx.$$
The operator $FF^T$ is also called the \emph{frame operator} of $\{f_i\}_{i=1}^k$. 

It is easy to see that $\sum_{i=1}^k{| \langle x,f_i\rangle |}^2$ remains unchanged if we put negative sign on any vector in $F$ or  permute the vectors in $F$. 
%Moreover, applying an orthogonal transformation on $F$ preserves the matrix $FF^T$, which preserves the value of $\sum_{i}{| \langle x,f_i\rangle |}^2$.
%Thus, it is natural to come up with the following definition.
We will also allow the rotation of frame vectors. So we have the following definition on frame equivalence.

\begin{definition}
Two frames $F=\{f_i\}_{i=1}^k$ and $G=\{g_i\}_{i=1}^k$ of $\R^n$ are 
\begin{enumerate}
\item \emph{Type I equivalent} if there exists an orthogonal matrix $U$ such that  $g_i=Uf_i$ for all $i$.
\item \emph{Type II equivalent} if $\{f_i\}_{i=1}^k$ is a permutation of $\{g_i\}_{i=1}^k$.
\item \emph{Type III equivalent} if $ f_i = \pm g_i$ for every i.

\end{enumerate}

Finally, we say that two frames are \emph{equivalent} if they belong to the same equivalence class in the equivalence relation generated by these three equivalence relations.

\end{definition}

Such definition is not new and can also be found, for example, in \cite{HP04}. However, such equivalence relation as defined above is different than the equivalence relation that is often used.\footnote{More commonly frames $\{f_i\}$ and $\{g_i\}$ are called equivalent provided that there is an invertible operator $T$ such that $Tf_i = g_i$ for all $i$.}

\begin{example}
Let $f_i=(\cos\frac{2\pi}{3}i, \sin\frac{2\pi}{3}i), i=0,1,2$, and let $U$ be a $2\times 2$ orthogonal matrix. 

$F_1=\{f_1, f_2, f_3\}$ is type I equivalent to $F_2=\{Uf_1,Uf_2,Uf_3\}$.

$F_1=\{f_1, f_2, f_3\}$ is type II equivalent to $F_3=\{f_3,f_2,f_1\}$.

$F_1=\{f_1, f_2, f_3\}$ is type III equivalent to $F_4=\{f_1,-f_2,-f_3\}$.

$F_1=\{f_1, f_2, f_3\}$ is equivalent to $F_5=\{Uf_2, Uf_3,-Uf_1\}$.
\end{example}

Given two frames $F=\{f_i\}_{i=1}^k$ and $G=\{g_i\}_{i=1}^k$, how can we determine if they are essentially the same, i.e., equivalent? It is easy to note that a necessary condition for equivalence is the set $\{\|f_i\|\}_{i=1}^k$ is the same as the set $\{\|g_i\|\}_{i=1}^k$. We will rule out this easy case and only consider \emph{unit norm frames}, which are frames whose frame vectors are  all unit norm. We use $\Sb(n,k)$ to denote all the unit norm frames of $\R^n$ that has $k$ frame vectors.

\begin{problem}
Given two unit norm frames $F,G\in \Sb(n,k)$, how to efficiently tell whether these two frames are equivalent?
\end{problem}

As far as we can tell, this question has not been properly addressed, and is not a trivial one. An exhaustive checking is combinatorial in nature and therefore not tractable. If we are only considering Type II and III equivalence, then the frame operator $FF^T=\sum_{i=1}^k f_if_i^T$ is unchanged. Allowing orthogonal transformation adds complication because the new frame is $UF$ and consequently the new frame operator is $UF(UF)^T=UFF^TU^T$, which is different from $FF^T$ in general. If we only want to check whether $G=UF$(Type III equivalent), we can check whether $F^TF=G^TG$ holds. $F^TF$ is called the \emph{Gram matrix} of the frame $F$. The entries of the gram matrix are the pairwise inner products of the frame vectors, i.e., $(F^TF)_{i,j}=\langle f_i, f_j\rangle$.

On the other hand, $FF^T=GG^T$ implies that $G=FQ$ for some $k\times k$ orthogonal matrix $Q$, which does necessarily mean $F$ and $G$ are equivalent.

With the above discussion, a naive way of solving Problem 1 is to check whether two frames differ by an orthogonal transform after all possible permutations and sign changing. This is undoubtedly expensive and unsustainable. 
%This exhaustive algorithm is presented in Section \ref{sec:n} with some numerical experiments.
% if and only if $G=UF$

%In this paper, we will introduce a way to  analyze the equality of frames. We have a great method in $\R^2$ and we have proved the feasibility of this method, which is much faster than a regular methods. The one we discoveblack can also be used in  $\R^n$ when $n \geq 3$, but it still needs some techniques to prove. 

This paper aims to provide an effective way to solve Problem 1, which is the Inner Product Algorithm presented in Section \ref{sec:rn}. Section \ref{sec:r2} specifically addressed the $\R^2$ case where a second algorithm, the Angle algorithm is also presented.  Section \ref{sec:rn} discusses the general $\R^n$ case. The theoretical support is provided for the $\R^2$ case, and we leave it to future work for the general case. We demonstrate the effectiveness of proposed algorithm in Section \ref{sec:ne}.

Throughout this paper, we use the notion $[m]$ for the index set $\{1,2,\cdots,m\}$. Moreover, $\R^m_+=\{(x_1,x_2,\cdots, x_m): x_i\geq0, i\in[m]\}.$ Moreover, for a matrix $A$, $abs(A)$ is matrix obtained after taking absolute value of every entry of $A$.

\section{The Angle method in $\R^2$}\label{sec:r2}

This section focuses on frames of $\R^2$. In this simple case, vectors can be oriented nicely and we can use the angles between them. In order to express the ideas effectively, we need to present a few definitions and lemmas.

\begin{figure}[hbt]
\centering
\begin{tikzpicture}
\draw (0,0) circle(2);
\draw[black, ultra thick] (0,0)--({2*cos(0)}, {2*sin(0)}) node[right] {$f_1$};
\draw[black, ultra thick] (0,0)--({2*cos(30)}, {2*sin(30)}) node[right] {$f_2$};
\draw[black, ultra thick] (0,0)--({2*cos(65)}, {2*sin(65)}) node[above] {$f_3$};
%\draw[black, ultra thick] (0,0)--({2*cos(60)}, {2*sin(60)}) node[above] {$f_4$};
%\draw[black, ultra thick] (0,0)--({2*cos(77)}, {2*sin(77)}) node[above] {$f_5$};
%\draw[black, ultra thick] (0,0)--({2*cos(93)}, {2*sin(93)}) node[above] {$f_6$};
\draw[black, ultra thick] (0,0)--({2*cos(110)}, {2*sin(110)}) node[above] {$f_{k-1}$};
\draw[black, ultra thick] (0,0)--({2*cos(150)}, {2*sin(150)}) node[left] {$f_k$};
\draw (0.5,0.2) circle(0) node[right] {$\alpha_1$};
\draw (0.22,0.6) circle(0) node[right] {$\alpha_2$};
%\draw (0.5,0.85) circle(0) node[right] {$\alpha_3$};
%\draw (0.1,1.15) circle(0) node[right] {$\alpha_4$};
%\draw (-0.2,1.35) circle(0) node[right] {$\alpha_5$};
\draw (-0.3,0.8) circle(0) node[right] {$...$};
\draw (-1.3,0.75) circle(0) node[right] {$\alpha_{k-1}$};
\draw [blue,thick,domain=0:150] plot ({cos(\x)}, {sin(\x)});
\draw (0,1) circle(0) node[above] {\textcolor{blue}{$A$}};
%\draw[blue, <->] (1.732/2,1/2) arc(30:40:2);
\end{tikzpicture}
\caption{Cross angle}\label{fig:crossangle}
\end{figure}
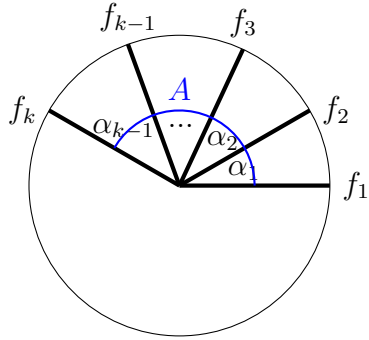

\begin{definition}
Given $F = \{f_i\}_{i=1}^k$, we can assume that $f_i$'s are oriented counterclockwise through Type II equivalence. Let $\alpha_i$ be the angle between $f_i$ and $f_{i+1}$, for any $i\in[k-1]$ (see Figure \ref{fig:crossangle}). Let $A=\sum_{i=1}^{k-1}\alpha_i$ and we can make $A<\pi$ through Type III equivalence. We call $A$ a \emph{cross angle} of $F$. Since rotation generates an equivalent frame,  $F$ can be characterized using the $k-1$ angles as $F\sim \{\alpha_i\}_{i=1}^{k-1}$. 
\end{definition}

%\textcolor{black}{Use all black in figures. Have figures centeblack like Figure 1. $f_6, f_5$ need to be aligned.}??

There are usually multiple cross angles associated with an equivalence class of frames. We list an example below.

\begin{figure}[htp]
\begin{subfigure}[t]{0.33\textwidth}
\begin{tikzpicture}
\draw (0,0) circle(2);
\draw[black, ultra thick] (0,0)--({2*cos(0)}, {2*sin(0)}) node[right] {$f_1$};
\draw[black, ultra thick] (0,0)--({2*cos(20)}, {2*sin(20)}) node[right] {$f_2$};
\draw[black, ultra thick] (0,0)--({2*cos(50)}, {2*sin(50)}) node[above] {$f_3$};
\draw[black, ultra thick] (0,0)--({2*cos(90)}, {2*sin(90)}) node[above] {$f_4$};
\draw[black, ultra thick] (0,0)--({2*cos(120)}, {2*sin(120)}) node[above] {$f_5$};
\draw[black, ultra thick] (0,0)--({2*cos(160)}, {2*sin(160)}) node[above] {$f_6$};

\draw (0.5,0.15) circle(0) node[right] {$\alpha_1$};
\draw (0.5,0.6) circle(0) node[right] {$\alpha_2$};
\draw (0,0.85) circle(0) node[right] {$\alpha_3$};
\draw (-0.7,1.05) circle(0) node[right] {$\alpha_4$};
\draw (-1.2,0.75) circle(0) node[right] {$\alpha_5$};
\draw [blue](0,1.25) circle(0) node[right] {$A_1$};
\draw [blue,thick,domain=0:160] plot ({cos(\x)}, {sin(\x)});
\end{tikzpicture}
\caption{Cross angle $A_1=\frac{8\pi}{9}$}
\label{fig:gull}
\end{subfigure}%
\begin{subfigure}[t]{0.33\textwidth}
\begin{tikzpicture}
\draw (0,0) circle(2);
\draw[black, ultra thick] (0,0)--({2*cos(0)}, {2*sin(0)}) node[right] {$f_1$};
\draw[black, ultra thick] (0,0)--({2*cos(20)}, {2*sin(20)}) node[right] {$f_2$};
\draw[black, ultra thick] (0,0)--({2*cos(50)}, {2*sin(50)}) node[above] {$f_3$};
\draw[black, dashed] (0,0)--({2*cos(90)}, {2*sin(90)}) node[above] {$f_4$};
\draw[black, dashed] (0,0)--({2*cos(120)}, {2*sin(120)}) node[above] {$f_5$};
\draw[black, dashed] (0,0)--({2*cos(160)}, {2*sin(160)}) node[above] {$f_6$};
\draw[ ultra thick] (0,0)--({2*cos(-60)}, {2*sin(-60)}) node[below] {$-f_5$};
\draw[ultra thick] (0,0)--({2*cos(-90)}, {2*sin(-90)}) node[below] {$-f_4$};
%\draw[ ultra thick] (0,0)--({2*cos(-130)}, {2*sin(-130)}) node[left] {$-f_3$};
\draw[ ultra thick] (0,0)--({2*cos(-20)}, {2*sin(-20)}) node[right] {$-f_6$};
\draw (0.64,0.15) circle(0) node[right] {$\alpha_1$};
\draw (0.56,0.6) circle(0) node[right] {$\alpha_2$};
\draw (0,0.85) circle(0) node[right] {$\alpha_3$};
\draw (-0.7,1.05) circle(0) node[right] {$\alpha_4$};
\draw (-1.2,0.75) circle(0) node[right] {$\alpha_5$};
\draw [blue](0.55,-0.9) circle(0) node[right] {$A_2$};
\draw [blue,thick,domain=-90:50] plot ({cos(\x)}, {sin(\x)});
\end{tikzpicture}
\caption{Cross Angle $A_2=\frac{7\pi}{9}$}
\label{fig:gull2}
\end{subfigure}%
        \begin{subfigure}[t]{0.33\textwidth}
\begin{tikzpicture}
\draw (0,0) circle(2);
\draw[black, ultra thick] (0,0)--({2*cos(0)}, {2*sin(0)}) node[right] {$f_1$};
\draw[black, ultra thick] (0,0)--({2*cos(20)}, {2*sin(20)}) node[right] {$f_2$};
\draw[black, ultra thick] (0,0)--({2*cos(50)}, {2*sin(50)}) node[above] {$f_3$};
\draw[black, ultra thick] (0,0)--({2*cos(90)}, {2*sin(90)}) node[above] {$f_4$};
\draw[black, ultra thick] (0,0)--({2*cos(120)}, {2*sin(120)}) node[above] {$f_5$};
\draw[black, dashed] (0,0)--({2*cos(160)}, {2*sin(160)}) node[above] {$f_6$};
\draw[ ultra thick] (0,0)--({2*cos(-20)}, {2*sin(-20)}) node[right] {$-f_6$};
%\draw[ ultra thick] (0,0)--({2*cos(-60)}, {2*sin(-60)}) node[below] {$-f_5$};
\draw (0.5,0.15) circle(0) node[right] {$\alpha_1$};
\draw (0.6,0.6) circle(0) node[right] {$\alpha_2$};
\draw (0,0.85) circle(0) node[right] {$\alpha_3$};
\draw (-0.7,1.05) circle(0) node[right] {$\alpha_4$};
\draw (-1.2,0.75) circle(0) node[right] {$\alpha_5$};
\draw [blue](1,0) circle(0) node[right] {$\beta_3$};
\draw [blue,thick,domain=-20:120] plot ({cos(\x)}, {sin(\x)});
\end{tikzpicture}
\caption{Cross Angle $A_3=\frac{7\pi}{9}$}
\label{fig:tiger}
\end{subfigure}%
\caption{Different cross angles for equivalent frames}\label{fig:minangle}
\end{figure}
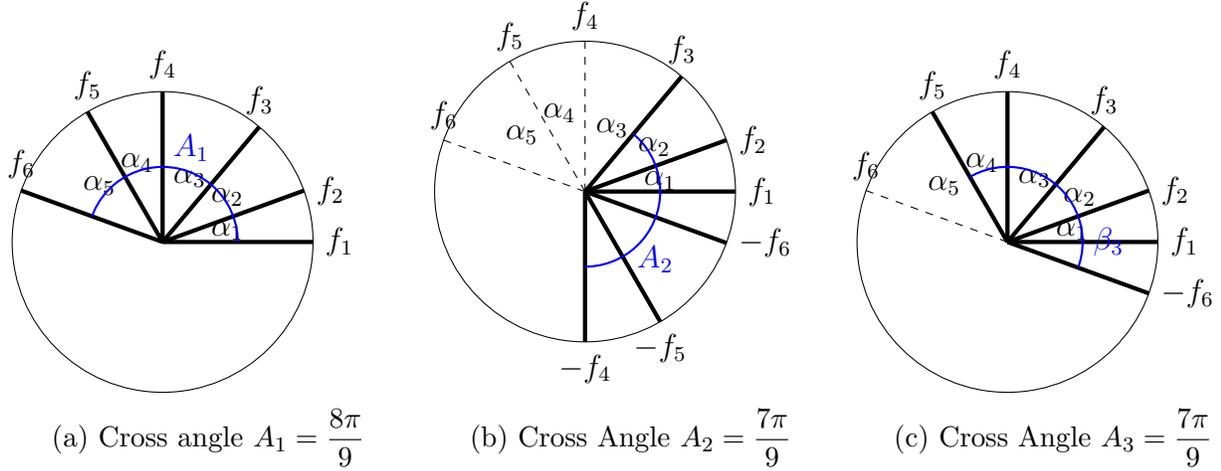

\begin{example}\label{ex:crossangle}
In Figure \ref{fig:minangle}(a), $\alpha_1 = \frac{\pi}{9}, \alpha_2 =\frac{\pi}{6}, \alpha_3 = \frac{2\pi}{9}, \alpha_4 = \frac{\pi}{6}, \alpha_5 = \frac{2\pi}{9}$. The cross angle is $A_1 = \frac{8\pi}{9}$.
If we pick the equivalent frame $\{-f_4, -f_5, f_1, f_2, f_3\}$ as shown in Figure \ref{fig:minangle}(b), then the cross angle is $A_2 = \frac{7\pi}{9}$.
Figure \ref{fig:minangle}(c) is $\{-f_6, f_1, f_2, f_3, f_4, f_5\}$ whose cross angle $A_3$ is also $\frac{7\pi}{9}$.
\end{example}

\begin{definition}
Given a frame $F$ of $\R^2$, define $A_F$ to be the smallest cross angle among all equivalent configurations of $F$.
\end{definition}

In Example \ref{ex:crossangle}, the smallest cross angle is $\frac{7\pi}{9}$, which is achieved in two different configurations (b) and (c). This is because they are both complement of $\alpha_3=\alpha_5$. In another word, the two pairs $f_3, f_4$, and $f_5, f_6$ produce the same angle $\frac{2\pi}{9}$. 

\begin{lemma}\label{lem:hard}
Suppose $F\sim \{\alpha_i\}_{i=1}^{k-1}$ has cross angle $A=\sum_{i=1}^{k-1}\alpha_i< \pi$, then all equivalent frames of $F$ has possible cross angles in the set  $\{A, \pi - \alpha_1, \pi - \alpha_2,...,\pi - \alpha_{k-1}\}$.
\end{lemma}
\begin{proof} 
Consider all vectors in $F$ and their negative ones, as showing in Figure \ref{fig:n}.
For convenience, name the angle between $f_{k}$ and $-f_1$ as $\alpha_{k}$.
Thus $A = \pi - \alpha_{k}$

\begin{figure}[htp]
\centering
\begin{tikzpicture}
\draw (0,0) circle(2);
\draw[black, ultra thick] (0,0)--({2*cos(0)}, {2*sin(0)}) node[right] {$f_1$};
\draw[black, ultra thick] (0,0)--({2*cos(20)}, {2*sin(20)}) node[right] {$f_2$};
\draw[black, ultra thick] (0,0)--({2*cos(35)}, {2*sin(35)}) node[above] {$f_3$};
\draw[black, ultra thick] (0,0)--({2*cos(60)}, {2*sin(60)}) node[above] {$f_4$};
\draw[black, ultra thick] (0,0)--({2*cos(77)}, {2*sin(77)}) node[above] {$f_5$};
\draw[black, ultra thick] (0,0)--({2*cos(93)}, {2*sin(93)}) node[above] {$f_6$};
\draw[black, ultra thick] (0,0)--({2*cos(130)}, {2*sin(130)}) node[above] {$f_{...}$};
\draw[black, ultra thick] (0,0)--({2*cos(150)}, {2*sin(150)}) node[above] {$f_k$};
\draw[blue, ultra thick] (0,0)--({-2*cos(0)}, {-2*sin(0)}) node[left] {$-f_1$};
\draw[blue, ultra thick] (0,0)--({-2*cos(20)}, {-2*sin(20)}) node[left] {$-f_2$};
\draw[blue, ultra thick] (0,0)--({-2*cos(35)}, {-2*sin(35)}) node[left] {$-f_3$};
\draw[blue, ultra thick] (0,0)--({-2*cos(60)}, {-2*sin(60)}) node[left] {$-f_4$};\draw[blue, ultra thick] (0,0)--({-2*cos(77)}, {-2*sin(77)}) node[below] {$-f_5$};
\draw[blue, ultra thick] (0,0)--({-2*cos(93)}, {-2*sin(93)}) node[below] {$-f_6$};
\draw[blue, ultra thick] (0,0)--({-2*cos(130)},{-2*sin(130)}) node[right] {$-f_{...}$};
\draw[blue, ultra thick] (0,0)--({-2*cos(150)}, {-2*sin(150)}) node[right] {$-f_k$};
\draw (0.5,0.15) circle(0) node[right] {$\alpha_1$};
\draw (0.6,0.45) circle(0) node[right] {$\alpha_2$};
\draw (0.5,0.85) circle(0) node[right] {$\alpha_3$};
\draw (0.1,1.15) circle(0) node[right] {$\alpha_4$};
\draw (-0.2,1.35) circle(0) node[right] {$\alpha_5$};
\draw (-1,1.35) circle(0) node[right] {$\alpha_{...}$};
\draw (-1.3,0.65) circle(0) node[right] {$\alpha_{k-1}$};
\draw [red](-1.6,0.3) circle(0) node[right] {$\alpha_{k}$};

%\draw[blue, <->] (1.732/2,1/2) arc(30:40:2);
\end{tikzpicture}
\caption{}\label{fig:n}
\end{figure}

It's clear that all angles between any two consecutive vectors are $\{\alpha_i\}_{i=1}^{k}$.

For any cross angle, it has two edge vectors, and the angle between the two edge vectors is exactly the cross angle. Also, all the other vectors in the frame are between the two edge vectors.
For example, the two edge vectors in Figure \ref{fig:minangle}(b) are $f_3$ and $-f_4$.

Let $g_n = \begin{cases}
      f_n & 1 \leq n\leq k \\
      -f_{n-k} & k+1 \leq n \leq 2k \\
\end{cases}$,
%and $ g_{2k+n} = g_{n}$
which denotes the sequence of vectors $\{\pm f_i\}$ counterclockwise as shown in Figure \ref{fig:n}. For convenience, we allow other indices of $g_n$ by making $g_n$ $2k$-periodic as $g_{2k+n}=g_n$.

%Clearly, it's easy to verify that given any vector $g_{i}$, its negative vector is $g_{i+k}$.

Let $g_i$ be an edge vector for some $i\in[2k]$. 
Consider the line formed by $g_i$ and $-g_i$. Since all cross angles are less than $\pi$, all the other vectors in the frame must be on one side of the line. %Otherwise the cross angle will be greater than $\pi$. 

Consider the two vectors $g_{i-1}$ and $g_{i+1}$ consecutive to $g_i$. Since $g_{i-1}$ and $g_{i+1}$ are on different sides of the line of $g_i$, one of them is in the frame, and the other one's negative vector is in the frame. 
Without loss of generality, assume $g_{i+1}$ is in the frame. 
Then this equivalent frame is $\{g_i, g_{i+1},g_{i+2}, \cdots, g_{i+k-1}\}$, making
 $g_{i+k-1}$ the other edge vector. 
 
 The cross angle is $\pi$ minus angle between $g_{i+k-1}$ and $g_{i+k}$. If $i\in[k]$, then angle between $g_{i+k-1}$ and $g_{i+k}$ is angle between $-f_{i-1}$ and $-f_{i}$, which is $\alpha_{i-1}$. If $k<i\leq 2k$, then angle between $g_{i+k-1}$ and $g_{i+k}$ is angle between $f_{i-k-1}$ and $f_{i-k}$, which is $\alpha_{i-k-1}$.
Thus the set of possible cross angles are $\{\pi - \alpha_{i}\}_{i=1}^k$.
\end{proof}

By Lemma \ref{lem:hard}, the smallest cross angle is the minimum of the set $\{A, \pi - \alpha_1, \pi - \alpha_2,...,\pi - \alpha_{k-1}\}$. If the minimum is achieved by only one element, then we have a unique minimal cross angle configuration which can be explicitly written down as shows in Lemma \ref{lem:hard}. This provides us an algorithm for determining equivalence of frames which we will call the Angle Algorithm.  

\vspace{0.1in}

\noindent\begin{tabular}{ll}
\multicolumn{2}{l}{\textbf{The Angle Algorithm: $\mathcal{O}(k\log k)$ }}\\
\hline
\hline
\multicolumn{2}{l}{Input: $F=\{f_i\}_{i=1}^k\in\Sb(2,k), G=\{g_i\}_{i=1}^k\in\Sb(2,k)$}\\
1: & Turn $F$ to its minimal angle form with angles $\alpha_1, \cdots, \alpha_{k-1}$: $\mathcal{O}(k\log k)$ \\
&1.1 Change signs of frame vectors of $F$ so that their $y$ coordinates are all nonnegative.\\
& Denote the new frame vectors as $f_1', \cdots, f_k'$ according to their $x$-values, descendingly. \\
& The angles between them are $\alpha_1', \cdots, \alpha_{k-1}'$\\
&1.2 $i_0=\arg\min  \{\langle f_i',f_{i+1}'\rangle\}_{i=1}^{k-1}$. This means $\alpha_{i_0}'$ is the biggest angle.\\
&1.3 The frame $\{-f_{i_0+1},-f_{i_0+2},\cdots, -f_{k}, f_1, f_2, \cdots, f_{i_0} \}$ is in its minimal angle form and \\
&equivalent to $F$.\\
2: & Turn $G$ to its minimal angle form with angles $\beta_1, \cdots, \beta_{k-1}$: $\mathcal{O}(k\log k)$ \\
3: & If $\alpha_i = \beta_i, \quad i=1,2,\cdots, k-1$ or $\alpha_i = \beta_{k-i}, \quad i=1,2,\cdots, k-1$, then they are equivalent.\\
\hline
\end{tabular}

\vspace{0.2in}

The rest of this section provides theoretical proof for this algorithm. The following Lemma will come handy in the main proof.

\begin{lemma}\label{lem:A}
Let $F\sim\{\alpha_i\}_{i=1}^{k-1}$ as defined earlier and 
 $A = \sum_{i=1}^{k-1}\alpha_i<\pi$ is a cross angle. $A$ is a minimal cross angle of $F$ if and only if $A + \max \{\alpha_i\} \leq \pi$.
\end{lemma}

\begin{proof}
$(\Longrightarrow)$ Assume that $A$ is a minimal cross angle. We  assume to the contrary that $A + \max \{\alpha_i\} > \pi$.
 then $A > \pi -  \max \{\alpha_i\} $. Let $\alpha_{i_0}=\max\{\alpha_i\}$, then we are able to generate an equivalent configuration whose cross angle is $\pi -  \max \{\alpha_{i_0}\}$ as shown in the proof of Lemma \ref{lem:hard}. This contradicts to $A$ being minimal.
%In this case, we can choose a smaller cross angle. For example, in Figure 2 - figureB, we choose to flip the angle $\alpha_2$, then we flip all the angles into that big angle we made(just like we did before), then we can get a new cross angle that is bigger than $\pi -max \{\alpha_i\}$, then we know that A is not the minimum. Therefore, we prove that  if A is minimum, then $A + \max \{\alpha_i\} \leq \pi$.\\

$(\Longleftarrow)$ Now we assume that  $A + \max \{\alpha_i\} \leq \pi$.
From Lemma \ref{lem:hard}, we know that $A, \pi - \alpha_1, \pi - \alpha_2,\cdots,\pi - \alpha_{k-1}$ are all the possible cross angles. $A + \max \{\alpha_i\} \leq \pi$ implies $A\leq\pi-\alpha_i$ for any $i\in[k-1]$, which shows that $A$ is a minimal cross angle.
\end{proof}

%With the definition of cross angles, we have an efficient way to determine whether two frames of $\R^2$ are equivalent due to the following theorem.

\begin{theorem}\label{thm:angle}
Let $F$ and $G$ be two frames of $\R^2$ that have unique minimal cross angle configuration. In their minimal cross angle form, let the angles of $F$ and $G$ be $\{\alpha_i\}_{i=1}^{k-1}$ and $\{\beta_i\}_{i=1}^{k-1}$ respectively. $F$ and $G$ are equivalent if and only if  one of the following two conditions is satisfied

(1) $\alpha_i = \beta_i, \quad i=1,2,\cdots, k-1$

(2)  $\alpha_i = \beta_{k-i}, \quad i=1,2,\cdots, k-1$.
\end{theorem}

\begin{proof}
With Type II (relabeling) and Type III equivalence, we can turn $F$ and $G$ to their corresponding minimal cross angle configurations as $F\sim\{\alpha_i\}_{i=1}^{k-1}$ and $G\sim\{\beta_i\}_{i=1}^{k-1}$.

Because the minimal configurations are unique, $F$ and $G$ are equivalent if and only if they differ by an orthogonal transformation, which must be a rotation or reflection. If it is a rotation, then $\alpha_i=\beta_i, i\in[k-1]$. If it is a reflection, then $\alpha_i = \beta_{k-i}, i\in[k-1]$.
\end{proof}

If there are multiple angles in  $\{A, \pi - \alpha_1, \pi - \alpha_2,...,\pi - \alpha_{k-1}\}$ (which happens with probability 0 if frames are random) achieving the minimal cross angles, the Angle algorithm will fail. This angle algorithm considers the generic case that there is only one minimal angle configuration (excluding frames in Example \ref{ex:crossangle}). The following section provides another algorithm that can deal with multiple minimal cross angle configuration case.

\section{Frames in $\R^n$: characterization by the gram matrix}\label{sec:rn}

If $n\geq3$, we can no longer resort to angles and the problem becomes a lot more complicated. In order to tackle this problem, we introduce a quantity that is invariant under any type of the equivalence transform.

\begin{definition}
Given $F=\{f_i\}_{i=1}^k\in\Sb(n,k)$ and $p>0$,  the quantity
\begin{equation}\label{equ:min}
FP_p(F)=\sum_{i<j}|\langle f_i,f_j\rangle|^p.
\end{equation}
is called the $p$-frame potential of F. 
\end{definition}

The concept of  frame potential was first introduced in \cite{BF03}, where $p=2$. The quantity $FP_2(F)$ can be used as a measure on how tight a frame is.
It was proven that the frames that achieve the smallest value of frame potential are precisely those that are tight. Later the work \cite{EO12} defines the general $p$-frame potential.

It is obvious that the $p$-frame potential remains the same under any equivalence transform, but is the opposite true? By \cite{BF03}, if $FP_2(F)=FP_2(G)$, then $F$ and $G$ are tight. But two tight frames are not necessarily equivalent. For example, let $F_0=\{(\cos \frac{i\pi}{4},\sin \frac{i\pi}{4})\}_{i=1}^4$ and $G_0=\{(1,0), (\cos\frac{\pi}{6}, \sin\frac{\pi}{6}), (0,1), (\cos\frac{2\pi}{3}, \sin\frac{2\pi}{3})\}$. We have that $$FP_2(F_0)=FP_2(G_0)=2,$$ but $F_0$ and $G_2$ are obviously not equivalent. On the other hand, it is true that $$FP_4(F_0)=1\neq \frac{5}{4}=FP_4(G_0).$$
This suggests that we may require the same frame potential energy for multiple values of $p$ to ensure equivalence. 

What is the consequence on requiring the same frame potential for multiple $p$? Given $x,y\in\R^m_{+}$, if $\sum_{i=1}^mx_i^p=\sum_{i=1}^my_i^p$ for enough powers $p$, one can imagine that $x$ must be a permutation of $y$. 

\begin{lemma}\label{lem:p}
Let $x,y\in\R^m_{+}$, if \begin{equation}\label{equ:p}
\sum_{i=1}^mx_i^p=\sum_{i=1}^my_i^p
\end{equation} for all $p>0$, then $x$ must be a permutation of $y$.
\end{lemma}
\begin{proof}
We need to prove that $x=y$ if both vectors are ordered ascendingly, that is $x_1\leq x_2\leq\cdots\leq x_m$ and $y_1\leq y_2\leq\cdots\leq y_m$.

We will prove by induction on $m$. The statement is obviously true for $m=1$. Now we assume it is true for $m-1$. In order to prove $x=y$, we first prove that $x_m=y_m$. Assume to the contrary that $x_m<y_m$, we divide \eqref{equ:p} by $y_m^p$, and get $$\sum_{i=1}^m\left(\frac{x_i}{y_m}\right)^p=\sum_{i=1}^{m-1}\left(\frac{y_i}{y_m}\right)^p+1$$
Taking limit of both sides as $p\rightarrow\infty$, we get $0=\lim_{p\rightarrow\infty}\sum_{i=1}^{m-1}\left(\frac{y_i}{y_m}\right)^p+1\geq1$, which is a contradiction.

This means that we must have $x_m=y_m$, which means $\sum_{i=1}^{m-1}x_i^p=\sum_{i=1}^{m-1}y_i^p$. By induction $x_i=y_i, i\in[m-1]$.
\end{proof}

Lemma \ref{lem:p} requires equal summation for all powers of $p$ (or at least infinitely many $p$'s going to infinity), but it is natural to think the result of Lemma \ref{lem:p} should still hold given \eqref{equ:p} is true for finitely many values of $p$. The following is such a result with vectors in $\R^3$.

\begin{lemma}\label{lem:3}
Let $x,y\in\R^3_+$. If
$$\sum_{i=1}^3x_i^p=\sum_{i=1}^3y_i^p, \text{ for }p=1,2,3,$$
then $x$ is a permutation of $y$.
\begin{proof}
Let $f_i: = x_1^i + x_2^i + x_3^i$ for $i=1,2,3$. We will first show that the numbers $x_1, x_2, x_3, y_1, y_2, y_3$ are all roots of a polynomial of degree 3.

Denote $e_2 =\sum_{i<j}x_ix_j$ and $e_3=x_1x_2x_3$. It is easy to see that 
\begin{equation}\label{equ:e2}
e_2 =  \frac{1}{2}(f_1^2-f_2).
\end{equation}
To compute $e_3$, we have
\begin{align*}
f_3 &= f_1f_2 - \sum_{i<j}(x_i^2x_j + x_j^2x_i)\\
&=f_1f_2 - \sum_{i<j}x_ix_j(f_1- x_l)\qquad (\{i,j,l\}=\{1,2,3\})\\
&= f_1f_2 - f_1e_2 +\sum_{i<j}x_ix_jx_l\\
&= f_1f_2 - f_1e_2 +3e_3.
\end{align*}
Thus,
\begin{equation}\label{equ:e3}
 e_3=\frac{1}{3}( f_3 - f_1f_2+f_1e_2)=\frac{1}{3}f_3-\frac{1}{2}f_1f_2 + \frac{1}{6}f_1^3.
 \end{equation}

On the other hand, we get $e_2 = x_ix_j + x_l(x_i+x_j) = x_ix_j+x_l(f_1-x_l)$, so
\begin{equation}\label{equ:e22}
x_ix_j = e_2-x_l(f_1-x_l)= \frac{1}{2}(f_1^2-f_2)-x_l(f_1-x_3)
\end{equation}

Compare \eqref{equ:e3} and \eqref{equ:e22}, we get that $x_l$  satisfies the following equation about $x$.
$$  \frac{1}{2}(f_1^2-f_2)x -(f_1-x)x^2  = \frac{1}{3}f_3-\frac{1}{2}f_1f_2 + \frac{1}{6}f_1^3$$

Since the above derivation also works for $y_i$, we have that  $x_1, x_2, x_3, y_1, y_2, y_3$ are all roots of the degree 3 polynomial $p(x)=  \frac{1}{2}(f_1^2-f_2)x -(f_1-x)x^2  -( \frac{1}{3}f_3-\frac{1}{2}f_1f_2 + \frac{1}{6}f_1^3)$. This implies that there are repeated roots among $\{x_1, x_2, x_3, y_1, y_2, y_3\}$.

If it is the case that $x_i=y_j$, then the arguments reduces to the dimension 2 case, which is true. Otherwise, we must have $x_1=x_2=x_3$ (or $y_1=y_2=y_3$), and consequently
\begin{align*}
&3x_1=y_1+y_2+y_3\\
&3x_1^2=y_1^2+y_2^2+y_3^2
\end{align*}
which forces $y_1=y_2=y_3$ and further that $x=y$.

\end{proof}
\end{lemma}

This suggests that the pairwise inner products (in absolute value) of $F$  should be a permutation of those of $G$. We know one direction is true.
\begin{lemma}\label{lem:one}
If $F,G\in\Sb(n,k)$ are equivalent, then the entries of $abs(F^TF)$ is a permutation of the entries of $abs(G^TG)$.
%$\{|f_i^Tf_j|, 1\leq i<j\leq k\}=\{|g_i^Tg_j|, 1\leq i<j\leq k\}$
\end{lemma}

\begin{proof}
With Type I equivalence, permutation does not affect the set of all inner products. With Type II equivalence, the absolute value absorbs all the sign effect. With Type III equivalence, inner products are preserved.
\end{proof}

\begin{theorem}\label{thm:3}
Let $F,G\in\Sb(2,3)$. $F$ and $G$ are equivalent if and only if 
\begin{equation}\label{equ:g}
\{|f_1^Tf_2|,|f_1^Tf_3|,|f_2^Tf_3|\}\text{ is a permutation of }\{|g_1^Tg_2|,|g_1^Tg_3|,|g_2^Tg_3|\}.
\end{equation}
\end{theorem}

\begin{proof}
One direction is covered by Lemma \ref{lem:one}. We need to prove the other direction and assume \eqref{equ:g} is true.

We can further  assume that both $F$ and $G$ are in their minimal angle configuration and $F\sim\{\alpha_1,\beta_1\}, G\sim\{\alpha_2,\beta_2\}$. We further assume that $\alpha_i\leq\beta_i$ through reflection. By Lemma \ref{lem:A}, $\alpha_i+\beta_i+\beta_i\leq\pi$, which implies that $\alpha_i\leq\beta_i<\frac{\pi}{2}$.

We also let $a_i=\cos\alpha_i, b_i=\cos\beta_i, c_i=\cos(\alpha_i+\beta_i)$. By the setup, $a_i\geq b_i>0$.

Lemma \ref{lem:3} implies that $\{a_1,b_1,|c_1|\}= \{a_2,b_2,|c_2|\}$. \eqref{equ:g} is equivalent to $\{a_1,b_1,|c_1|\}=\{a_2,b_2,|c_2|\}$.

Case 1: $c_1>0, c_2>0$. This would imply that $a_1\geq b_1>c_1, a_2\geq b_2<c_2$ so $a_1=a_2, b_1=b_2$, which directly implies $\alpha_1=\alpha_2, \beta_1=\beta_2$. So $F$ and $G$ are equivalent.

Case 2: $c_1<0, c_2<0$.

Case 2.1: additionally $a_1=a_2$. We assume $b_1=-c_2$ and $-c_1=b_2$, which implies $\alpha_1=\alpha_2$ and $\beta_1+\alpha_2+\beta_2=\pi$. In this case $F$ and $G$ are equivalent.

Case 2.2: additionally $a_1=b_2$. In this case $a_2\geq b_2=a_1\geq b_1$, so we can make $a_2=-c_1$ and $b_1=-c_2$, which implies that $\beta_2=\alpha_1, \beta_1+\beta_2+\alpha_2=\pi$. In this case $F$ and $G$ are equivalent.

Case 3: $c_1>0, c_2<0$. Since we already have $a_1\geq b_1>c_1, a_2\geq b_2$. We further split it to 3 cases:

Case 3.1: $a_2>b_2>-c_2$. This implies that $\alpha_1=\alpha_2, \beta_1=\beta_2$ and hence $c_1=c_2$. This case is not possible.

Case 3.2: $a_2\geq -c_2\geq b_2$. This means $a_1=a_2, b_1=-c_2, c_1=b_2$, implying $\alpha_1=\alpha_2, \beta_1+\alpha_2+\beta_2=\pi, \alpha_1+\beta_1=\beta_2$. This means that $\alpha_1+\beta_1=\frac{\pi}{2}$. $F$ and $G$ are equivalent in this case.

Case 3.3: $-c_2>a_2>b_2$. One has $\alpha_1+\alpha_2+\beta_2=\pi, \beta_1=\alpha_2, \alpha_1+\beta_1=\beta_2$. This implies $\alpha_1+\beta_1=\frac{\pi}{2}$ again. $F$ and $G$ are equivalent.

%Case 1.1: additionally $b_1=b_2$. This directly implies that $\alpha_1=\alpha_2, \beta_1=\beta_2$. So $F$ and $G$ are equivalent.
%
%Case 1.2: additionally $b_1=|c_2|, |c_1|=b_2$. In this case, if $c_1\geq 0$, then we have $\alpha_1+\beta_1=\beta_2$ and $\cos\beta_1=\cos(\alpha_2+\beta_2)=|\cos(2\alpha_1+\beta_1)|$, which implies that $\beta_1+2\alpha_1+\beta_1=\pi$. This means that $F\sim(\alpha_1,\frac{\pi}{2}-\alpha_1)$, and $G\sim(\alpha_1,\frac{\pi}{2})$. $F$ and $G$ are equivalent. If $c_1<0$
\end{proof}

\begin{theorem}
Let $F,G\in\Sb(2,3)$. $F$ and $G$ are equivalent if and only if 
\begin{equation}\label{equ:123}
FP_p(F)=FP_p(G),\text{ for }p=1,2,3
\end{equation}
\end{theorem}

\begin{proof}
Again one direction is clear. If \eqref{equ:123} holds, then \eqref{equ:g} holds by Lemma \ref{lem:3}. Therefore $F$ and $G$ are equivalent by Theorem  \ref{thm:3}.

\end{proof}

Now we present the general Inner Product Algorithm by simply comparing the gram matrix.

\vspace{0.1in}
\noindent\begin{tabular}{lll}
\multicolumn{3}{l}{\textbf{The Inner Product Algorithm: $O(\max\{\log k, n\}k^2)$}}\\
\hline
\hline
\multicolumn{3}{l}{Input: $F=\{f_i\}_{i=1}^k\in\Sb(n,k), G=\{g_i\}_{i=1}^k\in\Sb(n,k)$}\\
1:& Compute $\{|f_i^Tf_j|, 1\leq i<j\leq k\}$ and $\{|g_i^Tg_j|, 1\leq i<j\leq k\}$: &$(2n-1)k^2\text{ flops}$ \\ 
2: &sort $S=\{|f_i^Tf_j|, 1\leq i<j\leq k\}$&$2k^2\log(k)\text{ flops}$\\
3: &For any element $s\in\{|g_i^Tg_j|, 1\leq i<j\leq k\}$, check if $s\in S$ &$2k^2\log(k)$\\
&If they are the same, then $F$ and $G$ are equivalent, otherwise not equivalent.&\\
\hline
\end{tabular}

\vspace{0.2in}

The theoretical guarantee for $\Sb(2,3)$ is provided by Theorem \ref{thm:3}, and we list the following conjecture for the general case.
\begin{conjecture}
Let $F,G\in\Sb(n,k)$. $F$ and $G$ are equivalent if and only if $\{|f_i^Tf_j|, 1\leq i<j\leq k\}\text{ is a permutation of }\{|g_i^Tg_j|, 1\leq i<j\leq k\}$.
\end{conjecture}

%\begin{tabular}{lc}
%The Frame Algorithm \\
%\hline
%\hline
%%1: Procedure two frames G and F to know if they are equivalent or not.\\
%1: Given a frame F,  sum all inner products that are raised to the power p by using the equation: \\
%$\sum\limits_{i < j} {\langle f_i,f_j \rangle}^p$   total steps of that equation: $\dbinom{2}{k} -1 = \frac{k(k-1)}{2} -1 $\\
%for the inner product in dimension d by raising to the power p, total steps:\\
%$ (\frac{k(k-1)}{2} -1)* (2d -1 + 1 + p - 1) $\\
%2: sum all the equation m times to see if all the values are equal.\\
%total steps:  $\sum\limits_{p=1}^m (k+1)(k-2)(2d+p-1) =  (k+1)(k-2)(2md + \frac{m(m-1)}{2}-m) $\\
%$\qquad \qquad \qquad \qquad \qquad \qquad \qquad \qquad=  (k+1)(k-2)(2md + \frac{m^2}{2}-\frac{3m}{2})$\\
%3: If all the equations are equal, then we can say two frames are equivalent.\\
%\hline
%\end{tabular}\\
%
%\begin{align*}
%\text{total steps }&= \sum\limits_{p=1}^m (2d+p-1)(\frac{k(k-1)}{2}-1)*2\\
%		  & = \sum\limits_{p=1}^m (k(k-1) -2) (2d+p-1)\\
%		   &= \sum\limits_{p=1}^m (k+1)(k-2)(2d+p-1)\\
%		  & = (k+1)(k-2)(2md + \frac{m(m-1)}{2}-m)\\
%		  & =  (k+1)(k-2)(2md + \frac{m^2}{2}-\frac{3m}{2})\\
%\end{align*}
%
%
%\textcolor{red}{discuss $N=4, d=2$}
%\subsection{$\R^d, d\geq 3$}
%
%has to use frame potential (but didnt prove it...)

\section{Numerical experiments}\label{sec:ne}
\begin{figure}[htb]
\centering
\includegraphics[width=0.8\textwidth]{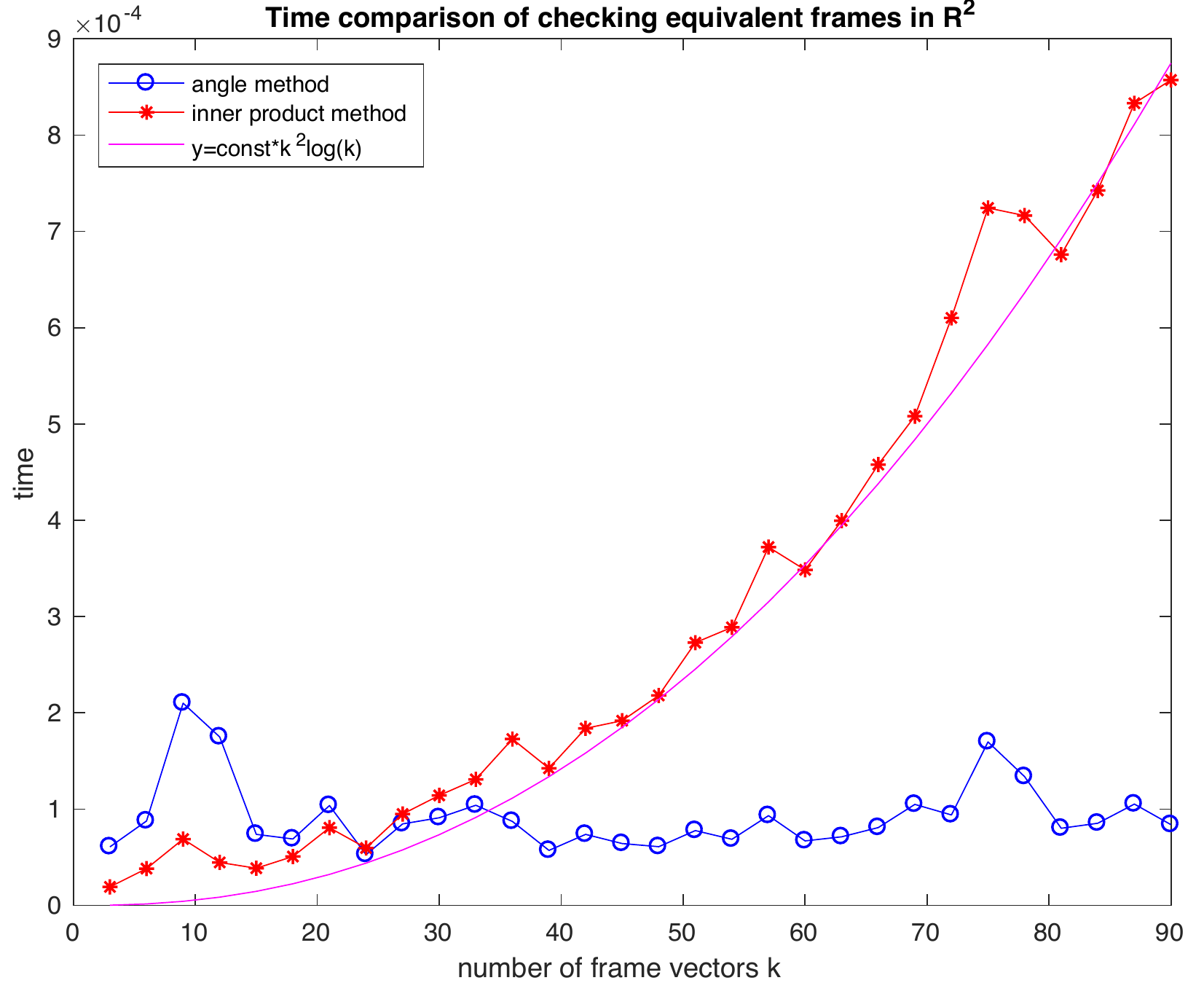}
\caption{Compare the angle algorithm and the inner product algorithm in $\R^2$ as the number of frame vectors $k$ grows. The inner product algorithm is exhibiting $k\log k$ growth.}\label{fig:r2}
\end{figure}

The first experiment is checking equivalence of frames  in $\R^2$ where both angle method and inner product method are used, as shown in Figure \ref{fig:r2}. For each $k$ ranging from 3 to 90 with an increment 3, we generate 20 random pairs of frames that contain both equivalent pairs and non-equivalent pairs. The graph is the time recorded after averaging over 20 pairs. One can see that the Angle Algorithm is more advantageous when $k\geq24$. In fact, the Angle method is exhibiting constant time over $k$ which is much better than expected. On the other hand, the time curve of the Inner Product Algorithm is showing $k\log k$ growth as computed.

\begin{figure}[htb]
\centering
\includegraphics[width=0.8\textwidth]{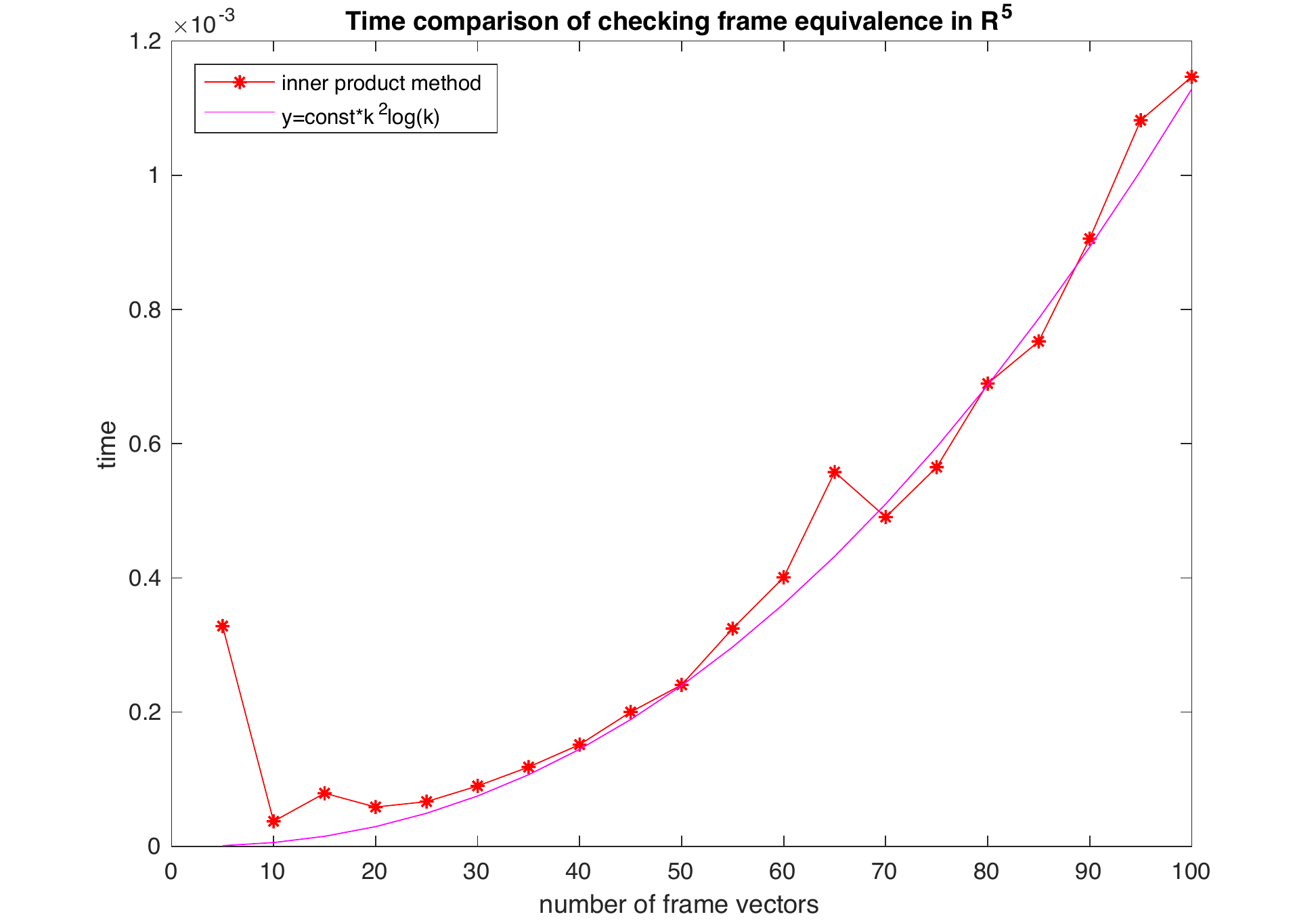}
\caption{Time used by the inner product algorithm in $\R^5$ as the number of frame vectors $k$ grows. The inner product algorithm is exhibiting $k\log k$ growth.}\label{fig:r5}
\end{figure}

The second experiment is in $\R^5$ and therefore only involves the Inner Product Algorithm. For each $k$ ranging from 5 to 100 with an increment 5, we generate 20 random pairs of frames that contain both equivalent pairs and non-equivalent pairs. The graph is the time recorded after averaging over 20 pairs. The complexity is showing a $k\log k$ growth again. See Figure \ref{fig:r5}.

\begin{figure}[htb]
\centering
\includegraphics[width=0.8\textwidth]{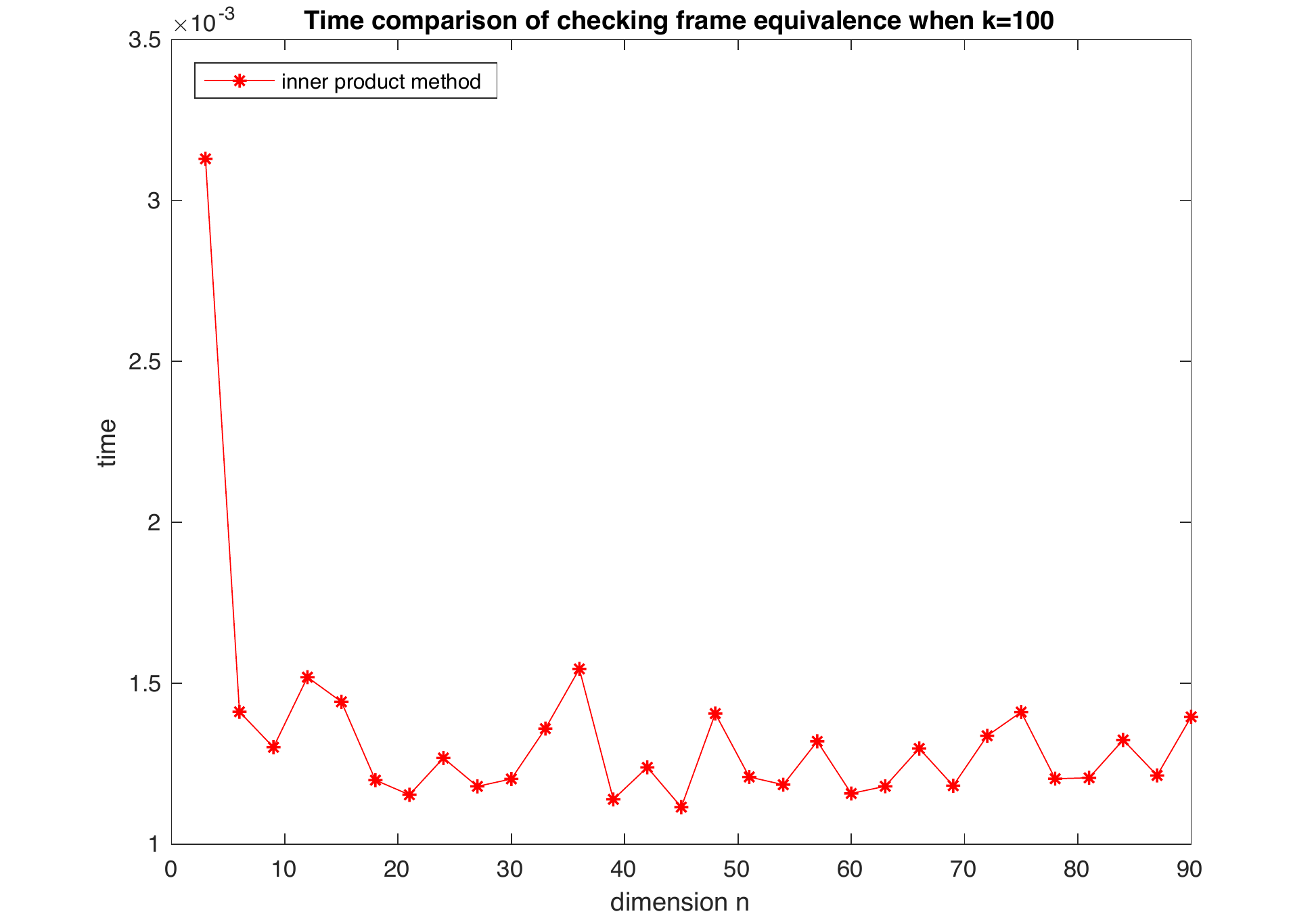}
\caption{Time used by the inner product algorithm in $\R^n$ as the dimension $n$ grows. The number of frame vectors $k$ is fixed to be 100.}\label{fig:k100}
\end{figure}

Figure \ref{fig:k100} displays the third experiment, which fixes the number of frame vectors to be $k=100$ and let the dimension $n$ grow from 3 to 90 with an increment of 3. The time spent is eventually constant over $n$, which is predicted.

\bibliographystyle{amsplain}

\end{document}